\documentclass[preprint,tightenlines,aps,floats,nofootinbib,showpacs]{revtex4}

\usepackage{amsmath,amssymb,array}
\usepackage{bm}
\usepackage{url}
\usepackage{graphicx}
\usepackage[dvips]{color}

\newcommand{\cbma}{\cos(\beta-\alpha)}
\newcommand{\sbma}{\sin(\beta-\alpha)}
\newcommand{\notE}{\hbox{{$E$}\kern-.60em\hbox{/}}}
\newcommand{\notp}{\hbox{{$p$}\kern-.43em\hbox{/}}}
\def\D0{\mbox{D\O}}

\newcommand{\LSB}{\left [}
\newcommand{\RSB}{\right ]}

\newcommand{\BR}{\mathcal{B}}

\newcolumntype{L}[1]{>{\raggedright\let\newline\\\arraybackslash\hspace{0pt}}m{#1}}
\newcolumntype{C}[1]{>{\centering\let\newline\\\arraybackslash\hspace{0pt}}m{#1}}
\newcolumntype{R}[1]{>{\raggedleft\let\newline\\\arraybackslash\hspace{0pt}}m{#1}}

\includegraphics{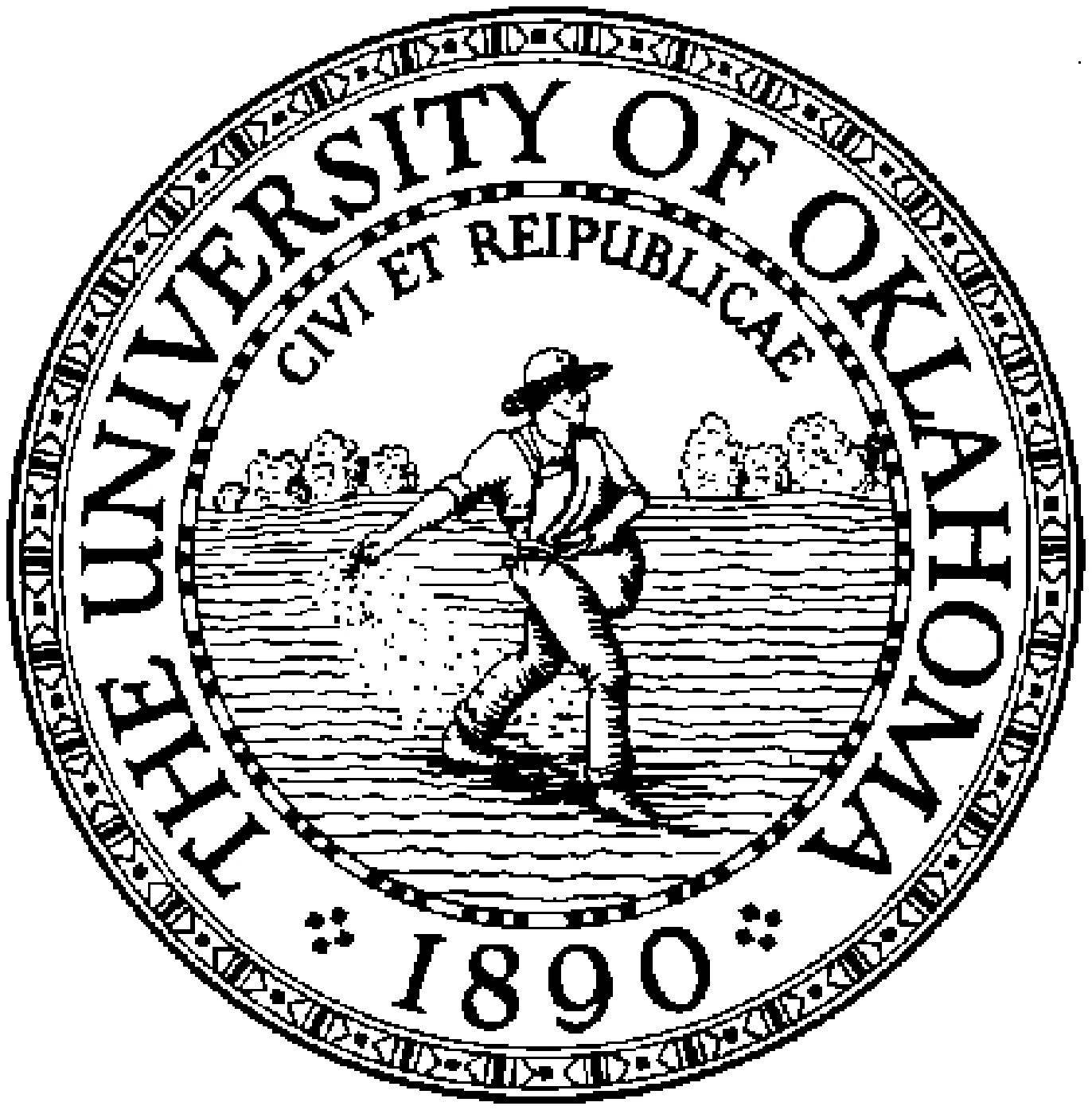}

\preprint{\font\fortssbx=cmssbx10 scaled \magstep2
\hbox to \hsize{
\hskip1.2in 
\hbox{\fortssbx The University of Oklahoma}
\hskip0.2in $\vcenter{
                      \hbox{\bf arXiv: [hep-ph]}
                      \hbox{\bf OU-HEP-150508}
                      \hbox{\bf CYCU-HEP-15-04}
                      \hbox{September 2015}}$ }
}

\begin{document}


\title{\vspace*{0.7in}
Flavor Changing Heavy Higgs Interactions at the LHC}

\author{
Baris Altunkaynak$^{a}$\footnote{E-mail address: baris@physics.ou.edu},
Wei-Shu Hou$^{b}$\footnote{E-mail address: wshou@phys.ntu.edu.tw},
Chung Kao$^{a}$\footnote{E-mail address: kao@physics.ou.edu},
Masaya Kohda$^{c}$\footnote{E-mail address: mkohda@cycu.edu.tw},
Brent McCoy$^{a}$\footnote{E-mail address: mccoy@physics.ou.edu}}

\affiliation{
$^a$Homer L. Dodge Department of Physics and Astronomy,
University of Oklahoma, Norman, OK 73019, USA \\
$^b$Department of Physics, National Taiwan University,
Taipei 10617, Taiwan, ROC\\
$^c$Department of Physics, Chung-Yuan Christian University,
Chung-Li 32023, Taiwan, ROC}

\date{\today}

\bigskip

\begin{abstract}
A general two Higgs doublet model (2HDM) is adopted to study the signature of
flavor changing neutral Higgs (FCNH) decay $\phi^0 \to t\bar{c}+\bar{t}c$, where
$\phi^0$ could be a CP-even scalar ($H^0$) or a CP-odd pseudoscalar ($A^0$).
%
%
Measurement of the light 125 GeV neutral Higgs boson ($h^0$) couplings at the
Large Hadron Collider (LHC) favor the decoupling limit or the alignment limit of
a 2HDM, in which gauge boson and diagonal fermion couplings of $h^0$ approach
Standard Model values. In such limit, FCNH couplings of $h^0$ are naturally
suppressed by a small mixing parameter $\cos(\beta-\alpha)$, while the
off-diagonal couplings of heavier neutral scalars $\phi^0$ are sustained by
$\sin(\beta-\alpha) \sim 1$.
We study physics background from dominant processes with realistic acceptance
cuts and tagging efficiencies. Promising results are found for the LHC running
at 13 or 14 TeV collision energies.
\end{abstract}

%


\maketitle

\newpage

\section{Introduction}

The Standard Model (SM) is very successful in explaining almost all experimental
data to date, culminating in the recent discovery of the long awaited Higgs
boson at the CERN Large Hadron Collider
(LHC)~\cite{ATLAS_Higgs_Discovery,CMS_Higgs_Discovery}. In the SM, all
elementary particles acquire mass from a single Higgs doublet that generates
spontaneous electroweak symmetry breaking (EWSB). All charged fermions have
their masses and Yukawa couplings to the Higgs boson as correlated but free
parameters. Furthermore, there are no flavor changing neutral currents (FCNC)
mediated by gauge interactions, nor by Higgs interactions (FCNH), at the tree
level. The most important goals of the LHC, at Run 2 and beyond, are the study
of Higgs properties and the search for signals, direct or indirect, of new
physics beyond the SM.

As the most massive particle ever discovered, the top quark might provide clues
to better understand the mechanism of EWSB. A possible explanation for its
heaviness could be provided by a special two Higgs doublet model for the top
quark (T2HDM)~\cite{Das:1995df}, where it is the only fermion that couples to a
Higgs doublet with a large vacuum expectation value (VEV). The second heaviest
particle is the newly discovered Higgs boson ($h^0$). With $m_{h^0} < m_t$, it
opens up the possibility of top quark decays into $h^0$ plus a charm quark. The
branching fraction of $t \to c h^0$ in SM at one loop level is approximately
$3 \times 10^{-15}$~\cite{Eilam:1990zc,Mele:1998ag,AguilarSaavedra:2004wm} 
for $m_{h^0} \simeq 120$ GeV. 
If this decay is detected, it would indicate a large effective FCNH
coupling of tree-level origins~\cite{Hou:1991un}, or very large enhancement from
beyond SM loop effects~\cite{AguilarSaavedra:2004wm}.

In flavor conserving two Higgs doublet models, a discrete
symmetry~\cite{Glashow:1976nt,Guide,Barger:1989fj} is often imposed to
distinguish the SU(2) doublet fields $\phi_1$ from $\phi_2$. Without such a
discrete symmetry, a general two Higgs doublet model (2HDM) should possess FCNH
vertices. To study such interactions, we adopt the following Lagrangian
involving Higgs bosons and fermions~\cite{Davidson:2005cw,Mahmoudi:2009zx}, 
\begin{eqnarray}
{\cal L}_Y &=& \frac{-1}{\sqrt{2}} \sum_{\scalebox{0.6}{F=U,D,L}}
 \bar{F}\left\{  \left[ \kappa^Fs_{\beta-\alpha}+\rho^Fc_{\beta-\alpha} \right] h^0 +
 \left[ \kappa^Fc_{\beta-\alpha}-\rho^Fs_{\beta-\alpha} \right] H^0 - i \, {\rm sgn}(Q_F)\rho^F A^0 \right\} P_R F \nonumber \\
 && -\bar{U} \left[ V \rho^D P_R - \rho^{U\dagger} V P_L \right] D H^+
  -\bar{\nu} \left[ \rho^L P_R \right] L H^+ + {\rm H.c.} \, ,
\end{eqnarray}
where $P_{L,R} \equiv ( 1\mp \gamma_5 )/2$,
$c_{\beta-\alpha} = \cos(\beta-\alpha)$,
$s_{\beta-\alpha} = \sin(\beta-\alpha)$,
$\tan\beta \equiv v_2/v_1$, and
$\alpha$ is the mixing angle between neutral Higgs scalars in the Type II
(2HDM-II) notation~\cite{Guide}. $\kappa$~matrices are diagonal and fixed by
fermion masses to $\kappa^F = \sqrt{2}m_F/v$ with $v \simeq 246$~GeV, while
$\rho$ matrices are free and have both diagonal and off-diagonal
elements. 
We adopt a CP conserving Higgs model and choose $\rho$ matrices 
to be real but not necessarily Hermitian.
$U$, $D$, $L$ and $\nu$ are vectors in flavor space ($U = (u,c,t)$, etc.). 
$h^0$ and $H^0$ are CP-even scalars ($m_h \leq m_H$), while $A^0$ is 
a CP-odd pseudoscalar.

With the advent of the LHC, theoretical interest in search of FCNH top decays
($t \to ch^0$) picked
up~\cite{Aguilar-Saavedra:2000aj,tch2011,tch2013,Atwood:2013ica}, and the ATLAS
and CMS experiments have already placed the branching fraction limit $\mathcal
B(t \to ch^0) < 5.6 \times 10^{-3}$~\cite{CMS_tch8}, implying
$\sqrt{\lambda_{htc}^2+\lambda_{hct}^2} < 0.14$. For LHC at $\sqrt{s} = 14$ TeV
and integrated luminosity of $L = 3000$ fb$^{-1}$, the ATLAS experiment
expects~\cite{ATLAS_tch14} to reach $\mathcal B(t \to ch^0) < 1.5 \times
10^{-4}$, i.e. probing down to
$\lambda_{htc} = \rho_{tc}\cos(\beta-\alpha)/\sqrt{2} < 0.024$.

The flavor changing heavy Higgs decay ($H^0 \to t\bar{c}+\bar{t}c$) is
complementary to FCNH top decay ($t \to ch^0$), since the coupling
$\lambda_{htc}$ is proportional to $\cos(\beta-\alpha)$ while $\lambda_{Htc}
\propto \sin(\beta-\alpha)$. Higgs boson data from the LHC favor the decoupling
limit~\cite{Gunion:2002zf} or the alignment
limit~\cite{Craig:2013hca,Carena:2013ooa} of a 2HDM. In this limit, FCNH
couplings of $h^0$ are naturally suppressed by small $\cos(\beta-\alpha)$, while
off-diagonal couplings of $H^0$, $A^0$ are sustained by $\sin(\beta-\alpha) \sim
1$.

In this letter, we study the discovery potential of the LHC in the search for
heavy Higgs bosons $H^0$ or $A^0$ that decay into a top quark and a charm quark.
The top quark then decays into a b quark, a charged lepton ($e$ or $\mu$), and a
neutrino. Taking LHC Higgs data and $B$ physics constraints into account, we
evaluate production rates with full tree-level matrix elements for both signal
and background. We optimize the acceptance cuts to effectively reduce the latter
with realistic $b$-tagging and mistag efficiencies. Promising results are
presented for the LHC with $\sqrt{s} = 13$ TeV as well as $14$ TeV.


\section{Constraints from Data}

In this section, we apply the latest results from LHC Higgs measurements, 
as well as from $B$ physics, to constrain the parameters $\rho_{tt}$,
$\rho_{bb}$, $\rho_{ct}$, $\rho_{tc}$, and $\cos(\beta-\alpha)$ of a general 2HDM
that are relevant for observing flavor changing decays of heavy Higgs bosons at the LHC.

\subsection{Constraints from ATLAS and CMS}

Run 1 of LHC at $\sqrt{s} = 7$ and 8 TeV has provided us with
information on the couplings of the Higgs boson $h^0$,
by measuring the event rates relative to the SM signal strength.
Even with our general 2HDM, the light Higgs boson $h^0$
constitutes a narrow resonance, and the signal strength for a
production channel $X$ and final state $Y$ can be written as
\begin{equation}
\mu_X(Y) =
 \frac{\sigma(X) \, \BR(h^0 \to Y)}{\sigma_{SM}(X)  \, \BR(h \to Y)_{SM}}.
\end{equation}

ATLAS and CMS often show the signal strength of measurements in two
dimensions by grouping gluon fusion and $tth$ production on one
axis (ggF), and vector boson fusion and associated
production on the other axis (VBF).
These contours can be used to draw constraints on
2HDM's~\cite{Belanger:2013xza}.
We follow a simpler approach and consider signal
strengths for final states with the largest statistics, namely
$\gamma \gamma$, $ZZ^*$, $WW^*$, and $\tau\tau$, where the dominant
production mode is gluon fusion, as well as the signal strength
for the $b\bar{b}$ final state from the associated production $Vh^0$
with $V = W$ or $Z$.
Table~\ref{tab:signalstrength} shows the average signal strengths
obtained by the experimental groups in Run 1.
We combine the ATLAS and CMS results and show both the combined 
values with their uncertainties in the last column.

\begin{table}[htb]
\renewcommand{\arraystretch}{1.7}
\begin{center}
\begin{tabular}{L{3.4cm}||C{2.3cm}|C{2.3cm}|C{2.3cm}}
Final state & $\mu$(ATLAS) & $\mu$(CMS) & $\mu$(comb.) \\
\hline \hline
$h^0 \to \gamma \gamma$   & $1.17^{+0.27}_{-0.27} \,$ \cite{Aad:2014eha} & $1.14_{-0.23}^{+0.26} \,$ \cite{Khachatryan:2014ira} & $1.16 \pm 0.18$ \\
$h^0 \to ZZ^* \to 4 \ell$ & $1.44^{+0.40}_{-0.33} \,$ \cite{Aad:2014eva} &  $0.93_{-0.25}^{+0.29} \,$ \cite{Chatrchyan:2013mxa} & $1.13 \pm 0.22$  \\
$h^0 \to WW^* \to \ell \nu \ell \nu$ & $1.09^{+0.23}_{-0.21} \,$ \cite{ATLAS:2014aga} &  $0.72_{-0.18}^{+0.20} \,$ \cite{Chatrchyan:2013iaa} & $0.89 \pm 0.14$\\
$h^0 \to \tau \tau$ & $1.43^{+0.43}_{-0.37} \,$ \cite{Aad:2015vsa} &  $0.78_{-0.27}^{+0.27} \,$ \cite{Chatrchyan:2014nva} & $0.99 \pm 0.22$ \\
$h^0 \to b \bar{b}$ & $0.52^{+0.40}_{-0.40} \,$ \cite{Aad:2014xzb} &  $1.00_{-0.50}^{+0.50} \,$ \cite{Chatrchyan:2013zna} & $0.71 \pm 0.31$
\end{tabular}
\caption{Signal strengths for the Higgs boson at the LHC.
The last column is our combination.
The combined signal strength for $h^0 \to WW^* +ZZ^* \, (VV)$ is
  $\mu(VV) = 0.96 \pm 0.12$.}
\label{tab:signalstrength}
\end{center}
\end{table}

To find the allowed regions of 2HDM parameter space 
compatible with ATLAS and CMS data, we take the discovered
Higgs boson as the lightest CP even state ($h^0$) of 
a general two Higgs doublet model and scan over the following sets of parameters:
\begin{equation}
\begin{split}
\textrm{General 2HDM:} & \quad \cos(\beta-\alpha) , \, \rho_{tt}, \,
 \rho_{bb}, \, \rho_{cc}, \, \rho_{\tau \tau} \, , \\
\textrm{Type II 2HDM:} & \quad \cos(\beta-\alpha), \, \tan\beta \, .
\end{split}
\label{eq:scanpars}
\end{equation}

In SM, the most important contributions to the Higgs total width 
come from $b\bar{b}$, $WW^*$, $gg$, $\tau\tau$, $c\bar{c}$ and $ZZ^*$.
The same channels are expected to contribute in a general 2HDM.
In our analysis, we include all the relevant parameters in
Eq.~(\ref{eq:scanpars}) that affect the total width,
gluon fusion cross section, and the associated Higgs production ($Vh^0$).

We cover a wide range of values for each free parameter,
and require that all Yukawa couplings of the mass eigenstates
($h^0$, $H^0$, $A^0$, $H^\pm$) stay perturbative, and that
the constraints given in Table~\ref{tab:signalstrength} are satisfied.
The signal strength for each production and decay channel can be
expressed in terms of scale factors which are couplings
of the Higgs boson to fermions and gauge bosons normalized to their
Standard Model values~\cite{LHCHiggsCrossSectionWorkingGroup:2012nn}.
These scale factors can then be expressed in terms of the
parameters of a specific 2HDM.

Negative results from heavy Higgs searches provide us further insights
on the parameter space of an extended Higgs sector. One of the strongest
results comes from the search for a heavy Higgs decaying into the $WW$
and $ZZ$ final states, excluding a heavy Higgs boson with SM like
couplings all the way up to 1 TeV~\cite{Khachatryan:2015cwa}. 
We use these results to further constraint the parameter space of 2HDM.

\begin{figure}[htb]
\begin{center}
\includegraphics[width=72mm]{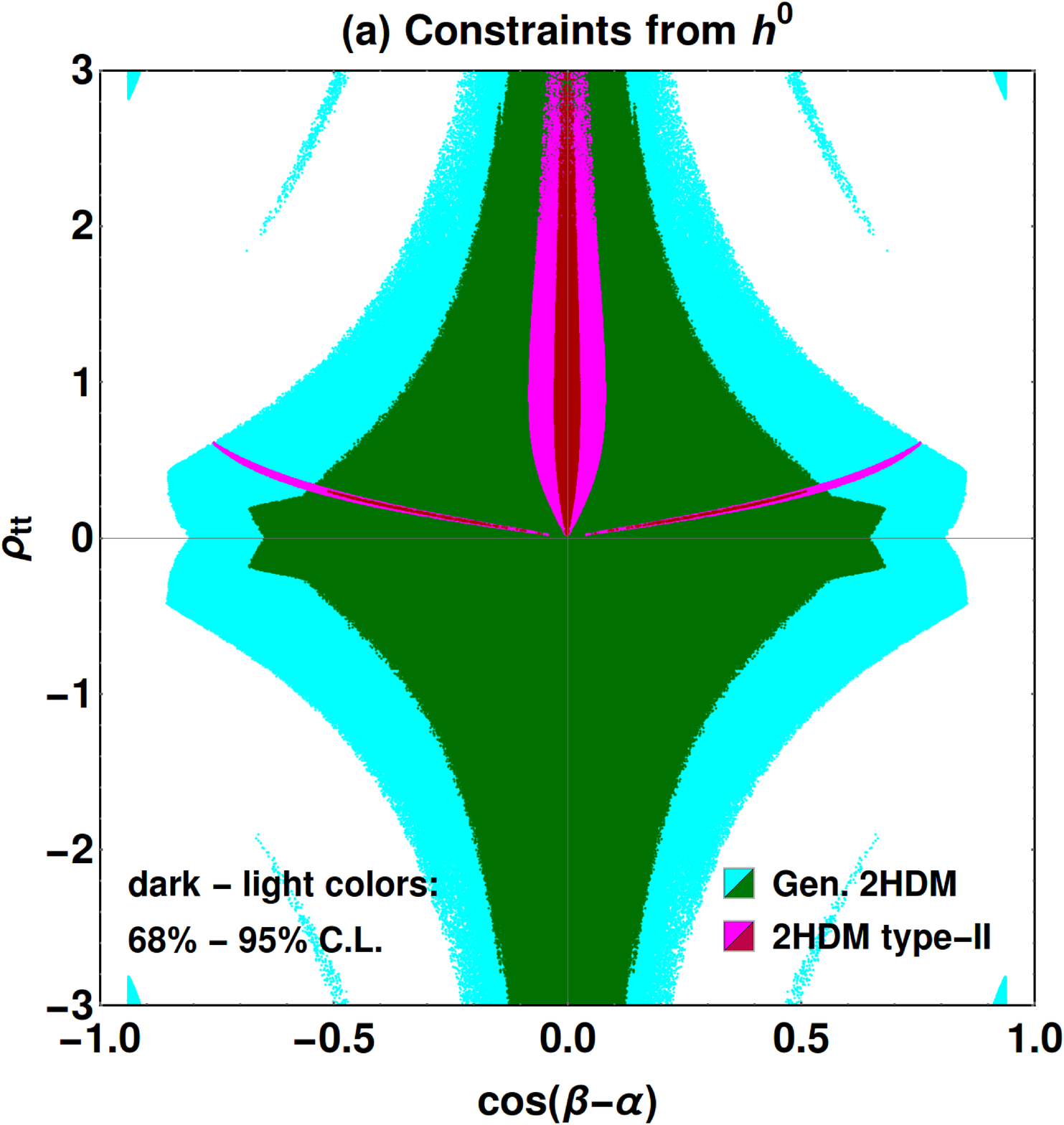}
\includegraphics[width=72mm]{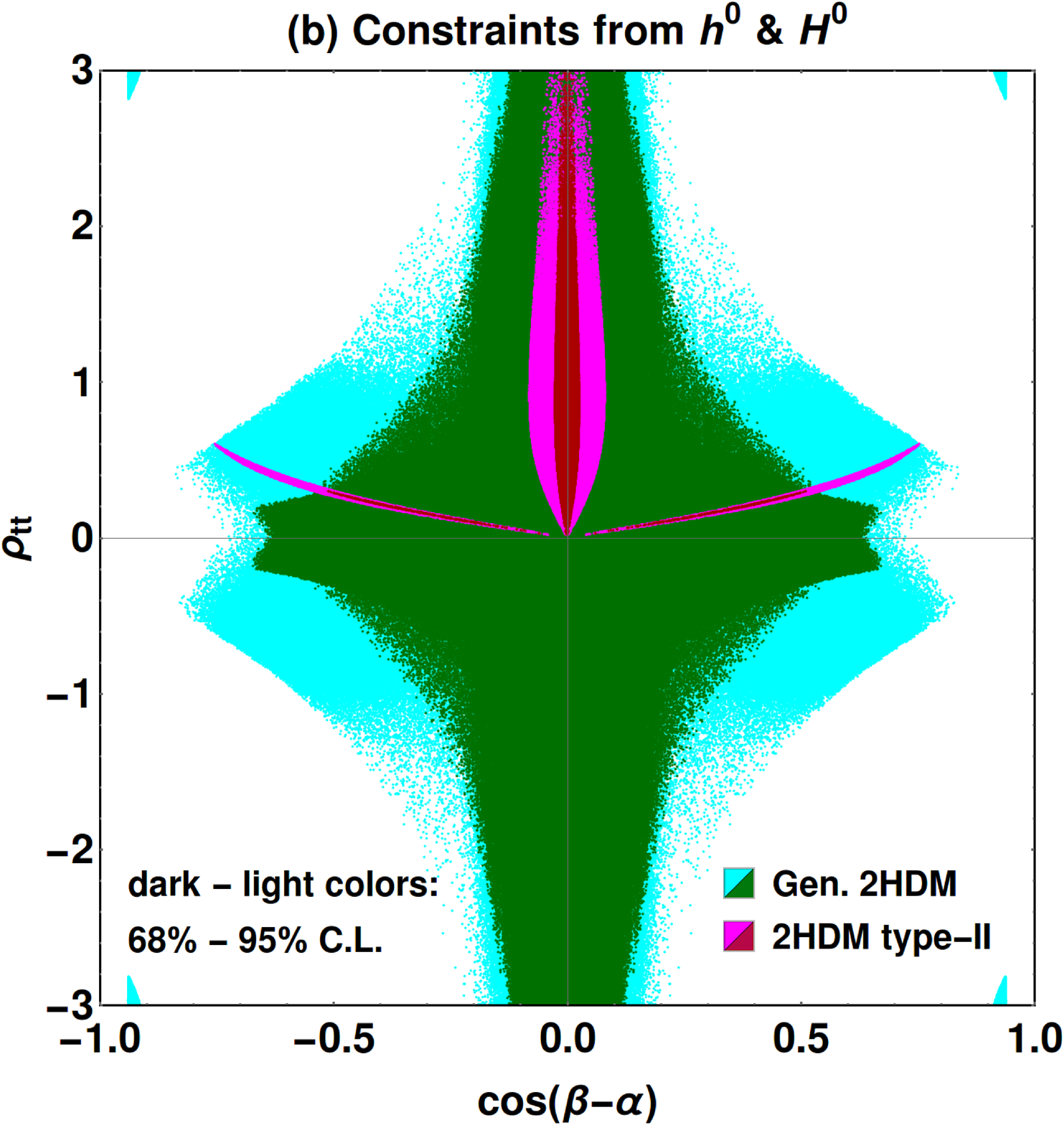}
\caption{
 Favored regions in $\cos(\beta-\alpha)$--$\rho_{tt}$ plane
 at 68\% (95\%) C.L. of LHC Higgs data 
 in a general 2HDM [green, (light cyan)] and 
 Type II 2HDM [red, (light magenta)], with constraints from 
(a) light Higgs boson ($h^0$), 
(b) both light Higgs boson ($h^0$) and heavy Higgs boson ($H^0$).
\label{fig:ATLAS_CMS}
}
\end{center}
\end{figure}

In Fig.~\ref{fig:ATLAS_CMS}, we present the 68\% (95\%) confidence
level (C.L.) regions in dark (light) color that are compatible with 
LHC constraints from the light Higgs boson ($h^0$) alone as well as 
constraints from both the light Higgs boson and the heavy Higgs boson
($H^0$) decaying into $WW$ and $ZZ$~\cite{Khachatryan:2015cwa} 
for a general 2HDM and and a Type-II 2HDM.
This figure shows that a large value of $|\cos(\beta-\alpha)|$
is still a possibility within a general model for $\rho_{tt} < 1$.
This is due to the lack of a strong constraint on the $b$-quark
coupling of the Higgs boson.
To be consistent with the SM Higgs cross section from gluon fusion,
a small value of $\cos(\beta-\alpha)$ is favored
for $\rho_{tt} \gg 1$ with
$\lambda_{htt} \sim \lambda_{htt}^{SM} = \kappa_{tt}$.
Experimental data from Run 2 with higher energy and higher luminosity
will provide much better guidance
for parameters such as $\rho_{tt}$ and $\cos(\beta-\alpha)$.


\subsection{Constraints from $B$ Physics}

The FCNH coupling $\rho_{ct}$ affects the $H^+tq$ couplings
($q=d,\,s,\,b$) through $(\rho^{U\dagger}V)_{tq} =
\rho_{tt}^*V_{tq}+\rho_{ct}^*V_{cq}+\rho_{ut}^*V_{uq}$.
This effect contributes to FCNC processes in down-type quark sector
via $H^+$ and $t$ loops.
For simplicity, let us assume $\rho_{ut}$ is negligible.

Recasting the 2HDM-II expression \cite{Geng:1988bq},
we estimate the modifications to the $B_q$-$\bar B_q$ ($q=d,s$) mixing
amplitude ($M_{12}^q$) from the box diagrams with internal $H^+/W$ and $t$ by
\begin{align}
\frac{M_{12}^q}{[M_{12}^q]_{\rm SM}} &=
 1+ \frac{I_{WH}(y^W,y^H,x) +I_{HH}(y^H)}{I_{WW}(y^W)}, \label{eq:B-mix}
\end{align}
where $y^i=m_t^2/m_i^2$ ($i=W,H^+$), $x=m_{H^+}^2/m_W^2$, and
\begin{align}
\begin{split}
&I_{WW}
 = 1+\frac{9}{1-y^W}-\frac{6}{(1-y^W)^2}
 -\frac{6}{y^W}\left(\frac{y^W}{1-y^W}\right)^3\ln y^W, \\
&I_{WH}
\simeq
 \left(\frac{\rho_{tt}^*}{\kappa_t}+\frac{V_{cb}\rho_{ct}^*}{V_{tb}\kappa_t}\right)
 \left(\frac{\rho_{tt}}{\kappa_t}+\frac{V_{cq}^*\rho_{ct}}{V_{tq}^*\kappa_t}\right)y^H \\
& \hspace{3cm} \times
  \left[ \frac{(2x-8)\ln y^H}{(1-x)(1-y^H)^2} +\frac{6x\ln y^W}{(1-x)(1-y^W)^2}
  -\frac{8-2y^W}{(1-y^W)(1-y^H)} \right], \\
&I_{HH}
\simeq
 \left(\frac{\rho_{tt}^*}{\kappa_t}+\frac{V_{cb}\rho_{ct}^*}{V_{tb}\kappa_t}\right)^2
 \left(\frac{\rho_{tt}}{\kappa_t}+\frac{V_{cq}^*\rho_{ct}}{V_{tq}^*\kappa_t}\right)^2 y^H
 \left[ \frac{1+y^H}{(1-y^H)^2}+\frac{2y^H\ln y^H}{(1-y^H)^3} \right].
\label{eq:B-mix-loop}
\end{split}
\end{align}
%
We adopt the following intervals from the Summer
2014 results by UTfit \cite{Bona:2006sa},
\begin{align}
C_{B_d}&\in [0.76,~1.43], \quad \phi_{B_d}\in [-8.0^\circ,~4.4^\circ], \notag\\
C_{B_s}&\in [0.9,~1.23], \quad \phi_{B_s}\in [-3.0^\circ,~4.7^\circ], 
\label{eq:UTfit}
\end{align}
at~95\%~probability,
where $C_{B_q}e^{2i\phi_{B_q}}\equiv M_{12}^q/[M_{12}^q]_{\rm SM}$. 

\begin{figure}[htb]
\begin{center}
\includegraphics[width=72mm]{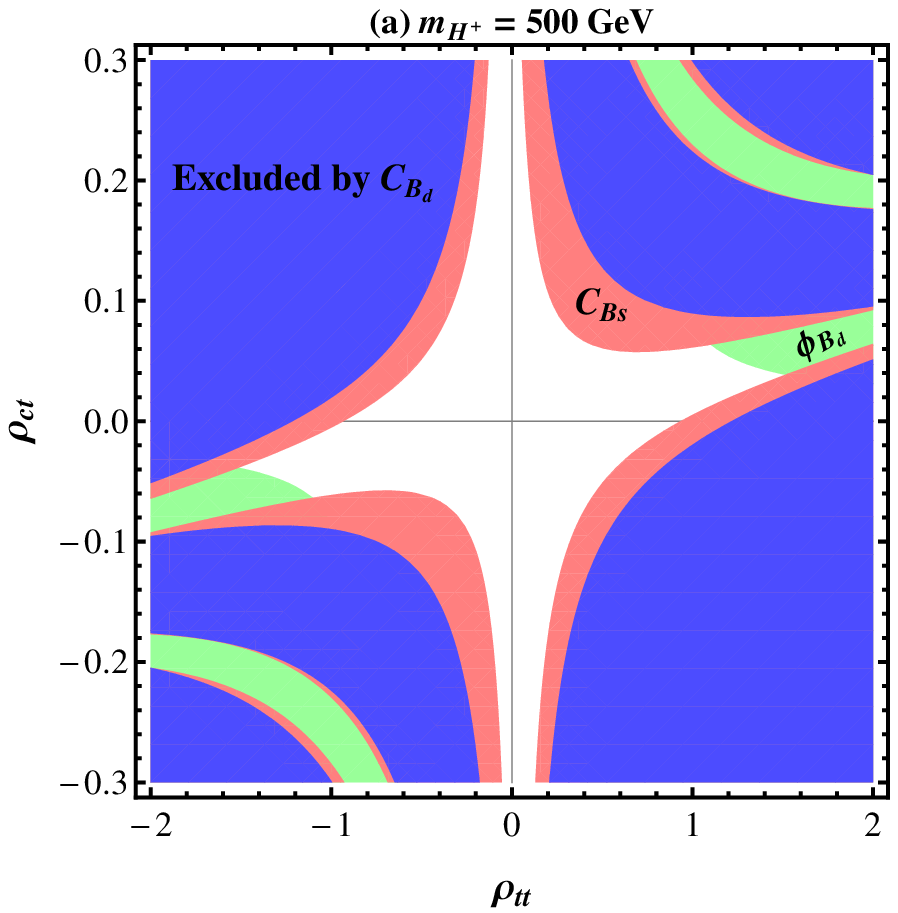}
\includegraphics[width=72mm]{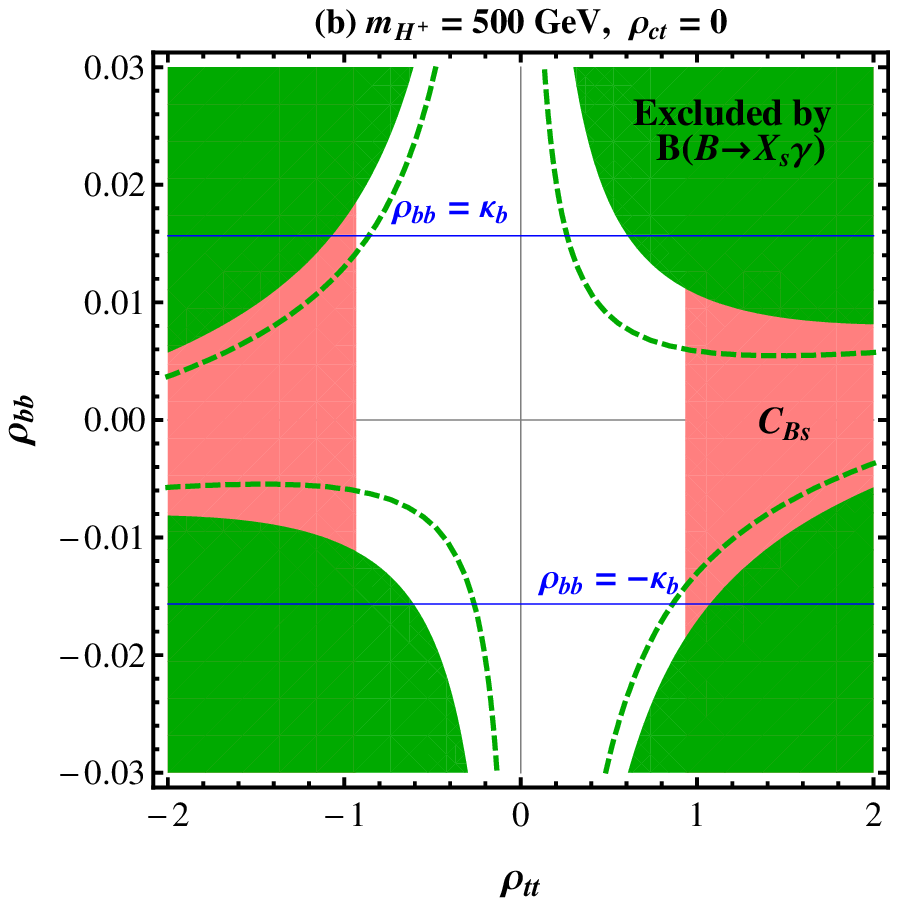}
\caption{
Allowed regions in
(a) $\rho_{tt}$--$\rho_{ct}$ plane from $B_{d,s}$-mixing, 
  with shaded regions excluded by Eq.~(\ref{eq:UTfit}),
  i.e. blue (pink) regions by $C_{B_{d(s)}}$, light-green regions by $\phi_{B_d}$;
(b) $\rho_{tt}$--$\rho_{bb}$ plane from $b \to s \gamma$
  with $\rho_{ct}=0$ and $m_{H^+}=500$ GeV,
  with dark-green shaded regions excluded
  by $R_{\rm exp}^{b\to s\gamma}$ in Eq.~(\ref{eq:b2sgam-exp}) at $2\sigma$,
  while pink regions by $C_{B_{s}}$.
Blue-solid lines in (b) mark $|\rho_{bb}|=\kappa_b$ with $m_b(\mu=m_t)$.
\label{fig:B_physics}
}
\end{center}
\end{figure}

The constraints from $B_{d,s}$ mixing data are shown in
Fig.~\ref{fig:B_physics}(a) on the $(\rho_{tt},\rho_{ct})$ plane
with $m_{H^+} = 500$ GeV.
Shaded regions are excluded by the 95\% probability ranges in
Eq.~(\ref{eq:UTfit}).
The constraint from $C_{B_s}$ (pink regions) is slightly
tighter than the $C_{B_d}$ exclusion (blue-shaded regions).
Combining them with constraint from the $CP$-violating phase
$\phi_{B_d}$ (light-green regions), we obtain the upper limit
$|\rho_{tt}|\lesssim 1.5$, regardless of $\rho_{ct}$.
The parameter $\rho_{ct}$ is strongly constrained since its effect
in Eq.~(\ref{eq:B-mix-loop}) is enhanced
by the CKM factor $|V_{cq}/V_{tq}| \sim 25$ ($q=d,s$).
Once $\rho_{tt}$ is fixed within this range, we obtain a constraint
on $\rho_{ct}$.
For $0.5 \alt |\rho_{tt}| \alt 1.5$, we have $|\rho_{ct}|\alt 0.06$.
Furthermore, the sizable phase in $V_{cd}/V_{td}$ makes $\rho_{ct}$ sensitive
to the $CP$-violating phase $\phi_{B_d}$, even if $\rho_{ct}$ is real.
For $m_{H^+}=300~(700)$ GeV, the constraints become:
$|\rho_{tt}|\lesssim 1.2~(1.8)$ regardless of $\rho_{ct}$, and
$|\rho_{ct}|\lesssim 0.05~(0.09)$
for $0.5\lesssim |\rho_{tt}|\lesssim 1.2~(1.8)$.



We now turn to the $b\to s\gamma$ constraint.
The $H^+$-$t$ loop affects this process via the Wilson coefficients $C_{7,8}$
at leading-order (LO), which are, at the matching scale $\mu_0$, given by
\begin{align}
\delta C_{7,8} \simeq
 \frac{1}{3} \left(\frac{\rho_{tt}^*}{\kappa_t} +\frac{V_{cb}\rho^{*}_{ct}}{V_{tb}\kappa_t}\right)
 \left(\frac{\rho_{tt}}{\kappa_t}+\frac{V_{cs}^*\rho_{ct}}{V_{ts}^*\kappa_t}\right)
  F_{7,8}^{(1)}(y^H)
 -\left(\frac{\rho_{tt}}{\kappa_t} +\frac{V_{cs}^*\rho_{ct}}{V_{ts}^*\kappa_t}\right)
  \frac{\rho_{bb}}{\kappa_b}F_{7,8}^{(2)}(y^H).
\label{eq:C7}
\end{align}
Here, the operator basis and the definition of $F_{7,8}^{(1,2)}(y)$ follow
Ref. \cite{Ciuchini:1997xe}.
We follow the procedure in Ref. \cite{Crivellin:2013wna}
and calculate first the ratio
\begin{align}
R^{b\to s\gamma}_{\rm exp}=
\frac{ \BR(B\to X_s \gamma)_{\rm exp}}{ \BR(B\to X_s \gamma)_{\rm SM}},
\label{eq:b2sgam-exp}
\end{align}
with world average of measurements $\BR(B\to X_s \gamma)_{\rm exp}
= (3.43\pm 0.21\pm 0.07)\times 10^{-4}$ \cite{Amhis:2014hma} and
the next-to-next-to LO (NNLO) prediction in SM, 
$\BR(B\to X_s \gamma)_{\rm SM} = (3.15\pm 0.23) \times 10^{-4}$ \cite{Misiak:2006zs}.
We then require the similar ratio based on our LO calculation, i.e.
$R^{b\to s\gamma}_{\rm theory}=\BR(B\to X_s \gamma)_{\rm 2HDM}/
\BR(B\to X_s \gamma)_{\rm SM}$,
to be within the range allowed by $R^{b\to s\gamma}_{\rm exp}$.
Choosing the matching scale and the low-energy scale
as $\mu_0 \sim m_t$ and $\mu_b \sim m_b/2$,
we reproduce the NNLO results of Ref.~\cite{Hermann:2012fc},
$m_{H^+} \geq 380$ GeV, with 95\% C.L.
in the Type-II 2HDM\footnote{
Recent NNLO calculation~\cite{Misiak:2015xwa}
provides a stronger limit ($m_{H^+}\geq 480$ GeV) at 95\% C.L.}.

Fig.~\ref{fig:B_physics}(b) shows the allowed region in the
$(\rho_{tt},\rho_{bb})$ plane from $b \to s\gamma$ and $B$ mixing for
$m_{H^+} = 500$ GeV and $\rho_{ct}=0$.
The dark-green shaded regions are excluded by the $2\sigma$
experimental error of $R_{\rm exp}^{b\to s\gamma}$
in Eq.~(\ref{eq:b2sgam-exp}),
with the theoretical uncertainty linearly added.
The constraint on $\rho_{tt}$ by $C_{B_s}$ is also shown.
For $\rho_{tt}\sim \kappa_t \sim 1$, $\rho_{bb}$ is constrained to be within
$-0.02\lesssim\rho_{bb}\lesssim 0.01$.
Note that this touches the region of $|\rho_{bb}| \sim \kappa_b\sim
0.02$.
We set $\rho_{ct}=0$ as it is already strongly
constrained by $B_{d,s}$ mixing.
For $m_{H^+}=300~(700)$ GeV, the $b \to s\gamma$ constraint on $\rho_{bb}$
becomes: $-0.009~(-0.03)\lesssim \rho_{bb}\lesssim 0.008~(0.02)$ for
$\rho_{tt}\sim 1$.
Typically, $\BR(B\to X_s\gamma)$ constrains $\rho_{bb}$ more strongly
than $\rho_{tt}$, as the effect of $\rho_{bb}$ is enhanced by the
chiral factor $\kappa_t/\kappa_b=m_t/m_b$  in Eq. (\ref{eq:C7}).

The contribution from charm loop has a mild dependence on $\rho_{tc}$
through the $H^+$ couplings, $(\rho^{U\dagger}V)_{cq} =
\rho_{tc}^*V_{tq}+\rho_{cc}^*V_{cq}+\rho_{uc}^*V_{uq}$ ($q=d,s,b$).
In general, $\rho_{tc}$ may be very different from $\rho_{ct}$.
Since the charm quark in the loop is light and there is no CKM
enhancement, the constraint on $\rho_{tc}$ is expected to be much weaker.
The constraint on $\rho_{tc}$ has been analyzed in
Ref.~\cite{Crivellin:2013wna}. With $-\epsilon_{32}^u = \rho_{tc} \sin\beta$,
it is found from $B_s$ mixing that $|\epsilon_{32}^u|\leq 1.7$
for $m_{H^+}=500$ GeV and $\tan\beta =50$.
This constraint implies that $|\rho_{tc}|\lesssim 1.7$ for $m_{H^+}=500$ GeV.
We note that a large $\rho_{tc}$ can enhance $B\to D^{(*)}\tau\nu$ rates
via the $H^+$ coupling $(\rho^{U\dagger}V)_{cb}$ and can explain
the 3.4$\sigma$ discrepancy between the SM prediction and
BaBar data~\cite{Lees:2012xj,Fajfer:2012jt,Crivellin:2012ye}.
For real valued $\rho_{tc}$ and $\rho_{\tau\tau}$,
the solution space shown in Ref.~\cite{Crivellin:2013wna} reads
$\rho_{tc}\sim 0.7\times(-0.5/\rho_{\tau\tau})(m_{H^+}/500~{\rm GeV})^2$.
Additional flavor constraints can be obtained from
$K-\bar{K}$ mixing ($\rho_{ct} \alt 0.14$)~\cite{Crivellin:2013wna} 
and $D-\bar{D}$ mixing ($|\rho_{tc}\rho_{tu}^*| \alt 0.02$)~\cite{
Crivellin:2013wna} for $m_H \simeq m_{H^+} = 500$ GeV.
The value of ($\rho_{tu}$) is expected to be very small, thus
$B-\bar{B}$ mixing provides a better limit for $\rho_{tc}$.


Combining experimental limits from LHC Higgs data and $B$ physics,
and assuming perturbativity, we consider
$\rho_{tt} < 2$, $\rho_{tc} < 1.5$, and $\rho_{ct} < 0.1$.

\section{Signal and Background}

We now discuss the prospects of discovering FCNH interactions at the LHC through
$H^0$ and $A^0$ decays. The number of free parameters in a general 2HDM is too
large for a comprehensive collider study of the FCNH signal, so we make some
assumptions that are motivated by experiment. The latest experimental results
point to a Higgs sector with the light CP even state behaving like the SM Higgs,
indicating that $\cos(\beta-\alpha)$ should be small. We consider sample cases
with $\cos(\beta-\alpha) = 0.1$ and $0.2$, which imply $\sin(\beta-\alpha) \sim
1$.

In our case study, we choose the heavier states ($H^0$, $A^0$ and $H^{\pm}$) to
be degenerate for simplicity, which is also in accordance with the decoupling
limit~\cite{Gunion:2002zf}, and we set $\lambda_{6,7} = 0$ in the Higgs
potential~\cite{Guide}. We also set $\tan\beta=1$ and choose $m_{12}^2$ such
that the scalar potential satisfies stability, tree-level unitarity, and
perturbativity up to large masses.

To fix the Yukawa couplings, we assume that $\rho_{tt} = \kappa_t$ while
$\rho_{bb} = \kappa_b$. This is in good agreement with both $B$ physics
constraints as well as LHC Run 1 constraints. For the off-diagonal parts that
generate the flavor changing signal, $\rho_{ct}$ is constrained to small values
by $B$~physics but we assume that $\rho_{tc}$ can have larger values. 
In the massless limit for charm, the signal cross section is, to a
very good approximation, only a function of a single effective coupling
$\tilde{\rho}_{tc} = \frac{1}{\sqrt{2}}\sqrt{\rho_{tc}^2 +\rho_{ct}^2}$, 
but also very weakly depends on 
$\frac{1}{\sqrt{2}}\sqrt{\rho_{tc}^2 - \rho_{ct}^2}$.
The contribution to the
cross section from terms with $\tilde{\rho}_{tc}$ is at least 98\% without cuts
and more than 93\% with all cuts for $pp \to H \to bc\ell\nu +X$ with $m_H = 1$
TeV, and it is even more dominating for a lower Higgs mass.

\begin{figure}[htb]
\begin{center}
\includegraphics[width=75mm]{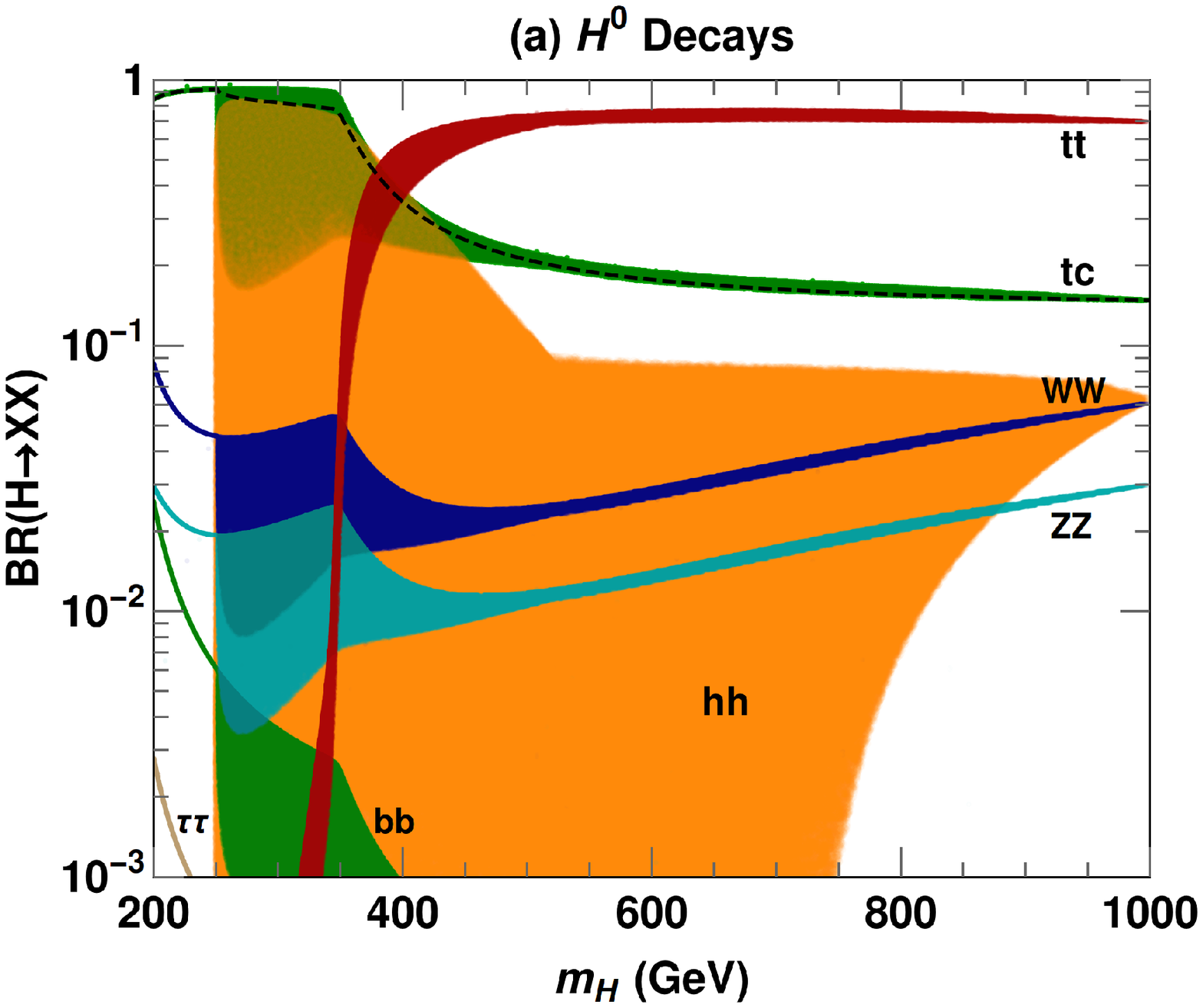}
\includegraphics[width=75mm]{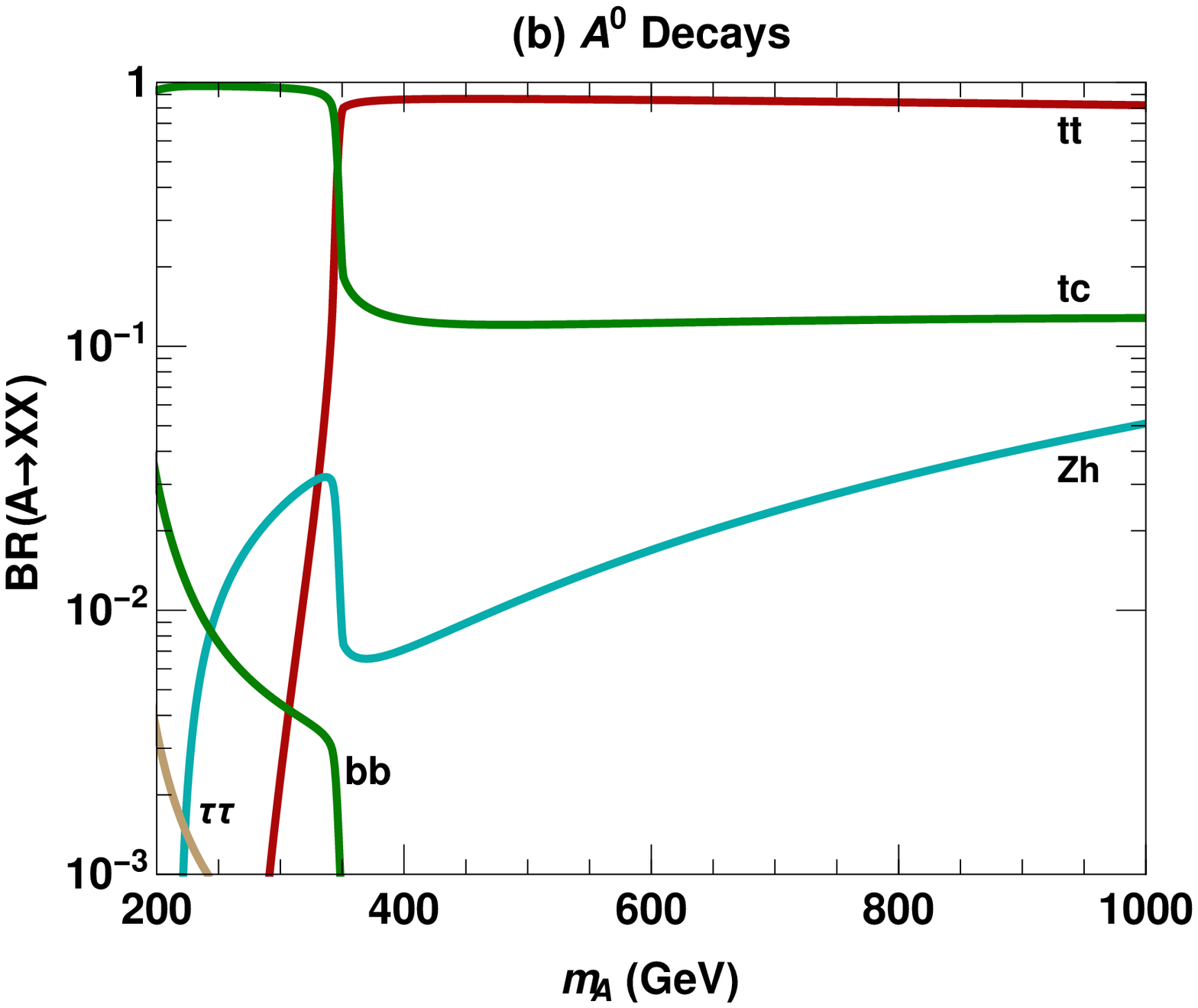}
\caption{
Branching fraction of (a) heavier Higgs scalar $H^0$
and (b) Higgs pseudoscalar $A^0$ versus $m_\phi$,
with $\cos(\beta-\alpha) = 0.1$, $\tilde{\rho}_{tc} = 0.24$,
and $\rho_{ii} = \kappa_i$ for diagonal couplings. We show the allowed regions when $\tan\beta$ and $m_{12}^2$ are varied. Branching fraction $\mathcal{B}(H^0 \to
tc)$ for the LHC case study is shown as a dashed curve.
\label{fig:BR13}
}
\end{center}
\end{figure}

With these experimentally motivated choices, gluon fusion is the dominant
production mode for $H^0$ and $A^0$ states, and $t\bar{t}$ becomes the dominant
final state at high mass ($2m_t < m_{\phi} \lesssim 2$ TeV). We display in
Fig.~\ref{fig:BR13} the branching fractions for (a) the heavier scalar $H^0$,
and (b) the pseudoscalar $A^0$, as functions of Higgs mass with
$\cos(\beta-\alpha) = 0.1$ and $\tilde{\rho}_{tc} = 0.24$.
The computer code
2HDMC~\cite{Eriksson:2009ws} is employed to scan over $|m_{12}| \le 2$ TeV and
$0.1 \le \tan\beta \le 50$ with sets of parameters that satisfy potential
stability, tree-level unitarity, and perturbativity. This gives rise to the
``bands'' in Fig.~3(a). We also display the branching fraction $\mathcal{B}(H^0 \to
tc)$ for the aforementioned choice of $\tan\beta$ and $m_{12}^2$ in our LHC
case study with a dashed curve in Fig.~3(a). We note that with large branching
fraction in most of the parameter space, $H^0 \to h^0 h^0$ might offer
great promise to discover Higgs pairs at the LHC.


\subsection{Higgs Signal}

Our signal is the production of a heavy Higgs boson from gluon fusion,
with subsequent flavor changing decays into a charm quark and a top quark,
and the top decays semileptonically.
More explicitly, we consider
$gg \to \phi^0 \to t\bar{c} +\bar{t}c$ ($\phi^0 = H^0$ or $A^0$),
followed by $t\bar{c}\to b\ell\nu \bar{c}$ with $\ell = e$ or $\mu$.
Unless explicitly specified,
$q$ generally denotes a quark ($q$) or an anti-quark ($\bar{q}$)
and $\ell$ represents a lepton ($\ell^-$) or anti-lepton ($\ell^+$).
We calculate the matrix elements analytically, 
and compute the signal cross section with the
parton distribution functions of MSTW2008~\cite{MSTW}.
The factorization and renormalization scales are
chosen to be $\mu_{F,R} = m_\phi$. In addition,
to estimate the NLO cross section for
$pp \to \phi^0 \to t\bar{c} +\bar{t}c \to b\ell\nu c +X$,
we use the computer code HIGLU~\cite{Spira:1995mt}
to calculate $\sigma(pp \to \phi^0 +X)$ ($\phi^0 = H^0, A^0$),
including both top and bottom quark loops to find a $K$-factor.

\subsection{Standard Model Background}

The dominant physics background to the final state of $bj\ell\nu$ comes from
$Wjj+Wb\bar{b}$, as well as $s$- and $t$-channel single top ($tb +tj$). Another
important background is $t\bar{t}$ production where either one of the two
leptons is missed for both top quarks decaying semileptonically, or two of the
four jets are missed when only one of top quarks decays semileptonically.
We employ the programs MadGraph~\cite{Stelzer:1994ta,Alwall:2007st} and
HELAS~\cite{Murayama:1992gi} to evaluate the exact matrix elements
for the background processes. The factorization and the
renormalization scales are chosen to be $\mu_{R,F} = m_W$ for $Wjj$
and $Wb\bar{b}$, $\mu_{R,F} = m_t$ for $s$- and $t$-channel single top, 
and $\mu_{R,F} = \sqrt{\hat{s}}$ for $t\bar{t}$.  We
use MCFM~\cite{Campbell:2010ff} to calculate the NLO $K$-factors for
our background processes.


\subsection{Mass Reconstruction}

Let us present our strategy for full reconstruction of
each event with the help of intermediate on-shell particles.
For each event, we require one $b$ jet and one non-$b$ jet,
identified through $b$-tagging. In addition, we require
a single isolated lepton and missing transverse energy
from the neutrino in the semileptonic decay
of the top quark in our FCNH signal.
For lepton momentum $p$ and neutrino momentum $k$,
the invariant mass constraint for an on-shell $W$, $(k+p)^2 = m_W^2$,
can be solved for the longitudinal component of the neutrino momentum
($k_z$), which is the only unknown in the event.
We obtain two solutions
\begin{eqnarray}
k_z^{\pm} =  \frac{p_z (2 {\mathbf{k}_T}\cdot {\mathbf{p}_T} + m_W^2 - m_\ell^2) \pm
 E_\ell \Delta }{2(m_\ell^2 + p_T^2)} \, ,
\Delta^2  =
  (2 {\mathbf{k}_T}\cdot {\mathbf{p}_T} +m_W^2 -m_\ell^2)^2 -4k_T^2 (m_\ell^2+p_T^2) \, .
\end{eqnarray}
If $\Delta^2 < 0$ hence $k_z^{\pm}$ complex, the event is vetoed.
For $\Delta^2 > 0$ with $k_z^{\pm}$ real,
we choose the solution that minimizes the
reconstructed top mass $|M_{b\ell\nu}-m_t|$ or $|(p_b+p+k)^2-m_t^2|$.

Systematics can be the limiting factor for new physics searches
at high luminosities. Precise determination of the background needs to
include systematics in experiments. Since our signal is a sharp peak over a
smoothly falling background, the precise knowledge of background cross
section at percent level is not required for a $5\sigma$ discovery.
An uncertainty of 30\% in the background estimation might shift the
limit on $g_{Htc}$ by 10\% without affecting our results.

\subsection{Realistic Acceptance Cuts}

To study the discovery potential, 
we employ three sets of realistic cuts and tagging efficiencies.
For low luminosity (LL):
 (a) LHC Run~1 ($\sqrt{s} = 8$ TeV) with $L =$25 fb$^{-1}$~\cite{LHC_Lumi}; and
 (b) full CM energy ($\sqrt{s} = 13$ or 14 TeV) with $L = 30$ fb$^{-1}$.
For high luminosity (HL):
 (c) full CM energy ($\sqrt{s} = 14$ TeV) with $L = 300$ fb$^{-1}$
     or $3000$ fb$^{-1}$\cite{ATLAS,CMS}.

We require that in every event there should be
(a) exactly 2 jets that have $p_T > 20$ GeV (30 GeV for HL) and $|\eta| < 2.5$,
and one of them must be tagged as a $b$-jet;
(b) exactly one isolated lepton that has $p_T > 20$ GeV and $|\eta| < 2.5$;
(c) at least 20 GeV (40 GeV for HL) of missing transverse energy ($\notE_T$).
After reconstructing the longitudinal component of the neutrino
momentum, we further require that
(d) the reconstructed invariant mass of the top satisfies
$|m_{bl\nu} - m_t| < 0.2 \, m_t$;
(e) the reconstructed invariant mass of the Higgs boson satisfies
$|m_{bl\nu c} - m_{\phi}| < 0.2 \, m_{\phi}$.

We consider a further powerful acceptance cut on non-$b$-tagged jet momentum. 
In the Higgs boson decay frame,
the charm quark momentum from $H^0,A^0 \to tc$ is approximately given by
\begin{equation}
p_c 
 \approx \frac{m_\phi}{2} \LSB 1 - \frac{m_t^2}{m_\phi^2} \RSB.
\end{equation}
Since the Higgs boson from gluon fusion has little transverse momentum,
the $p_T(c)$ distribution has both a kinematic cut-off
and a peak at the above $p_c$ value. 
We require that the transverse momentum of the non-$b$-tagged jet
satisfies $ 0.85 \, p_c < p_T(c) < 1.10 \, p_c$.

To simulate detector effects based on ATLAS~\cite{ATLAS}
and CMS~\cite{CMS} specifications, we apply Gaussian smearing of momenta:
\begin{eqnarray}
\frac{\Delta E}{E} = \frac{0.60}{\sqrt{E({\rm GeV})}} \oplus 0.03
\;\; ({\rm jets})\, , \quad {\rm and} \quad
\frac{\Delta E}{E} = \frac{0.25}{\sqrt{E({\rm GeV})}} \oplus 0.01
\;\; ({\rm leptons})\, ,
\end{eqnarray}
with individual terms added in quadrature ($\oplus$).
In LHC Run~1 at $\sqrt{s} = 8$ TeV, the $b$-tagging
efficiency ($\epsilon_b$) is taken to be $50\%$, the probability that
a $c$-jet is mistagged as a $b$-jet ($\epsilon_c$) is $14\%$ and the
probability that any other jet is mistagged as a $b$-jet
($\epsilon_j$) is taken to be $1\%$.
At the full CM energy ($\sqrt{s} = 13$ or 14 TeV) with $L=$ 30 fb$^{-1}$, 
we follow the tagging and mistag efficiencies 
in the ATLAS Technical Design Report~\cite{ATLAS}:
$\epsilon_b = 60\%$, $\epsilon_c = 14\%$ and $\epsilon_j = 1\%$.
For the full CM energy ($\sqrt{s} = 13$ or 14 TeV)
with HL of 300 fb$^{-1}$ or 3000 fb$^{-1}$,
the tagging and mistag efficiencies are taken to be
$\epsilon_b = 50\%$, $\epsilon_c = 14\%$ and $\epsilon_j = 1\%$.


\begin{figure}[htb]
\begin{center}
\includegraphics[width=72mm]{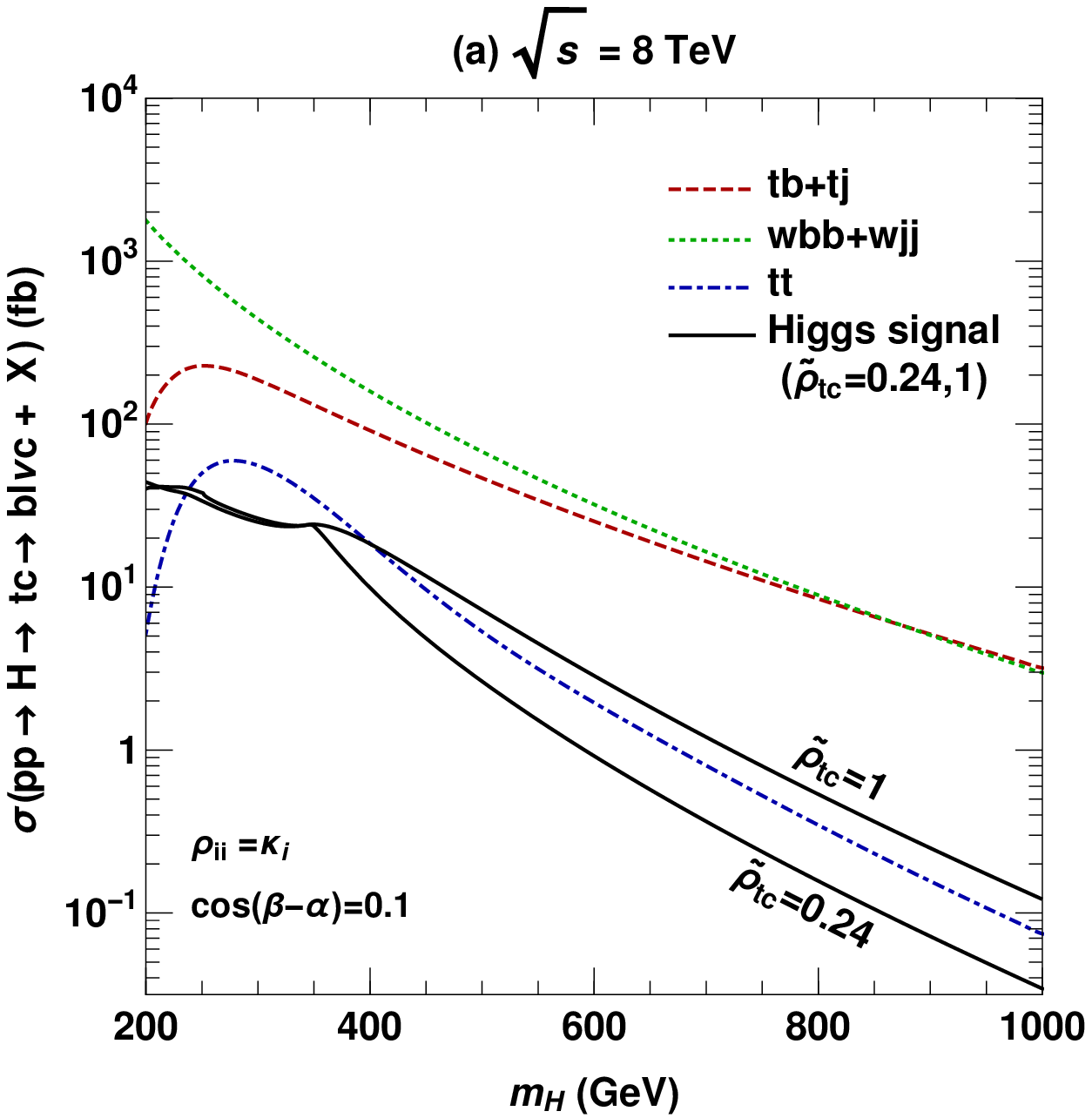}
\includegraphics[width=72mm]{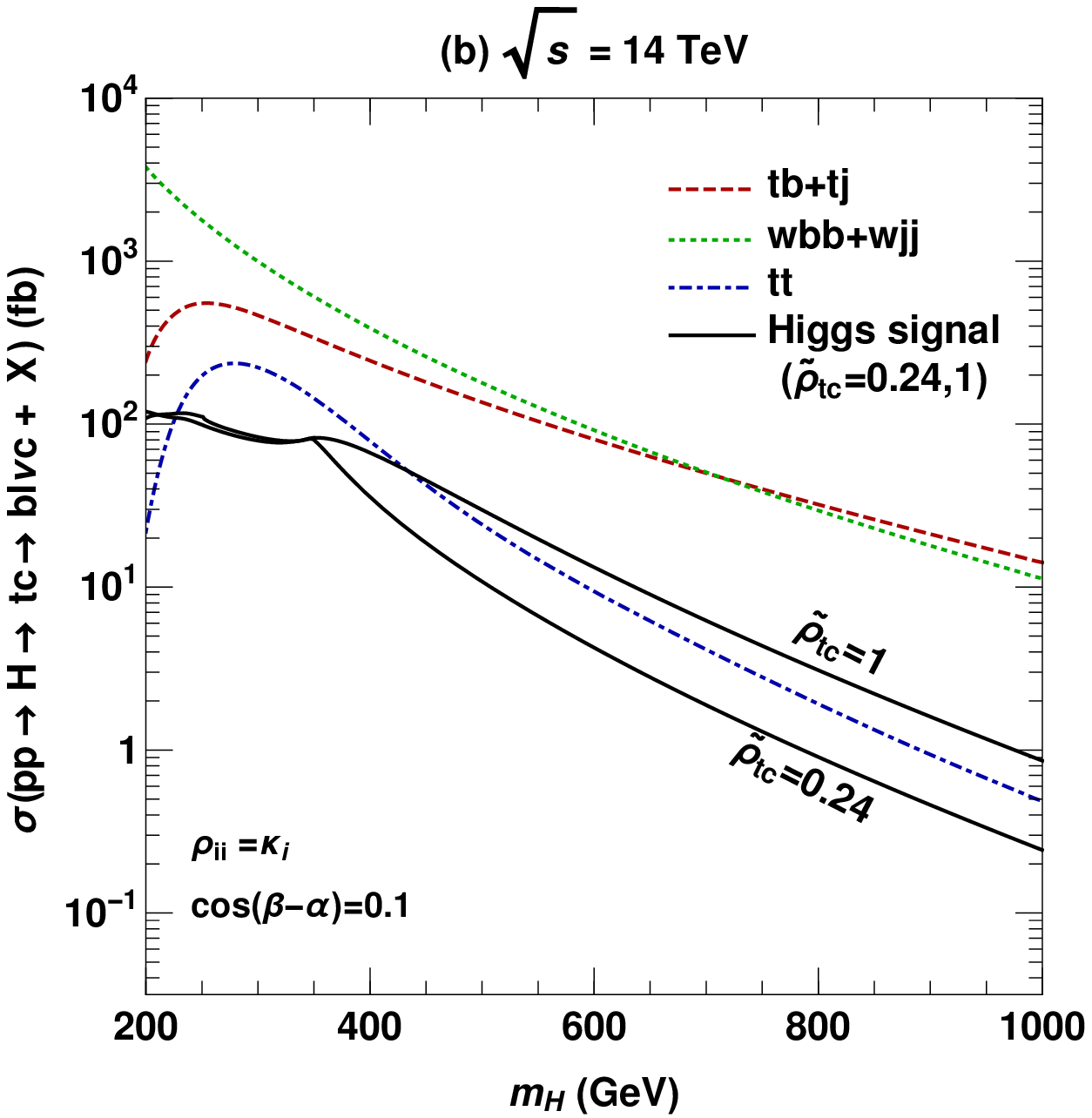}
\caption{
The cross section of the heavier Higgs scalar $H^0$ (solid)
$\sigma(pp \to H^0 \to t \bar{c} +\bar{t}c \to bj\ell +\notE_T +X)$
at the LHC versus $m_H$ for
 (a) $\sqrt{s} = 8$ TeV and
 (b) $\sqrt{s} = 14$ TeV,
with $\tilde{\rho}_{tc} = 0.24,\; 1$ and $\cbma = 0.1$.
Also shown are the background cross sections (dashed) from
single top ($tb$ and $tj$), $W$+jets ($Wjj$ and $Wbb$) and $t\bar{t}$
with $K$-factors, acceptance cuts, and tagging efficiencies.
\label{fig:sigma}
}
\end{center}
\end{figure}

\section{Discovery Potential} 

We present the signal and background cross sections at the LHC
for $\sqrt{s} = 8$ TeV and $14$ TeV in Fig.~\ref{fig:sigma}.
All tagging efficiencies and $K$-factors discussed above are included.
We observe that the largest contributions to the
SM background come from single-top and $W$+jets processes,
which is to be expected, since they can both produce
very similar kinematics to our signal process.
In contrast, the $t\bar{t}$ background is substantially lower
because of the requirement on the number of jets and leptons passing our cuts.


\begin{figure}[htb]
\begin{center}
\begin{tabular}{cc}
\includegraphics[width=66mm]{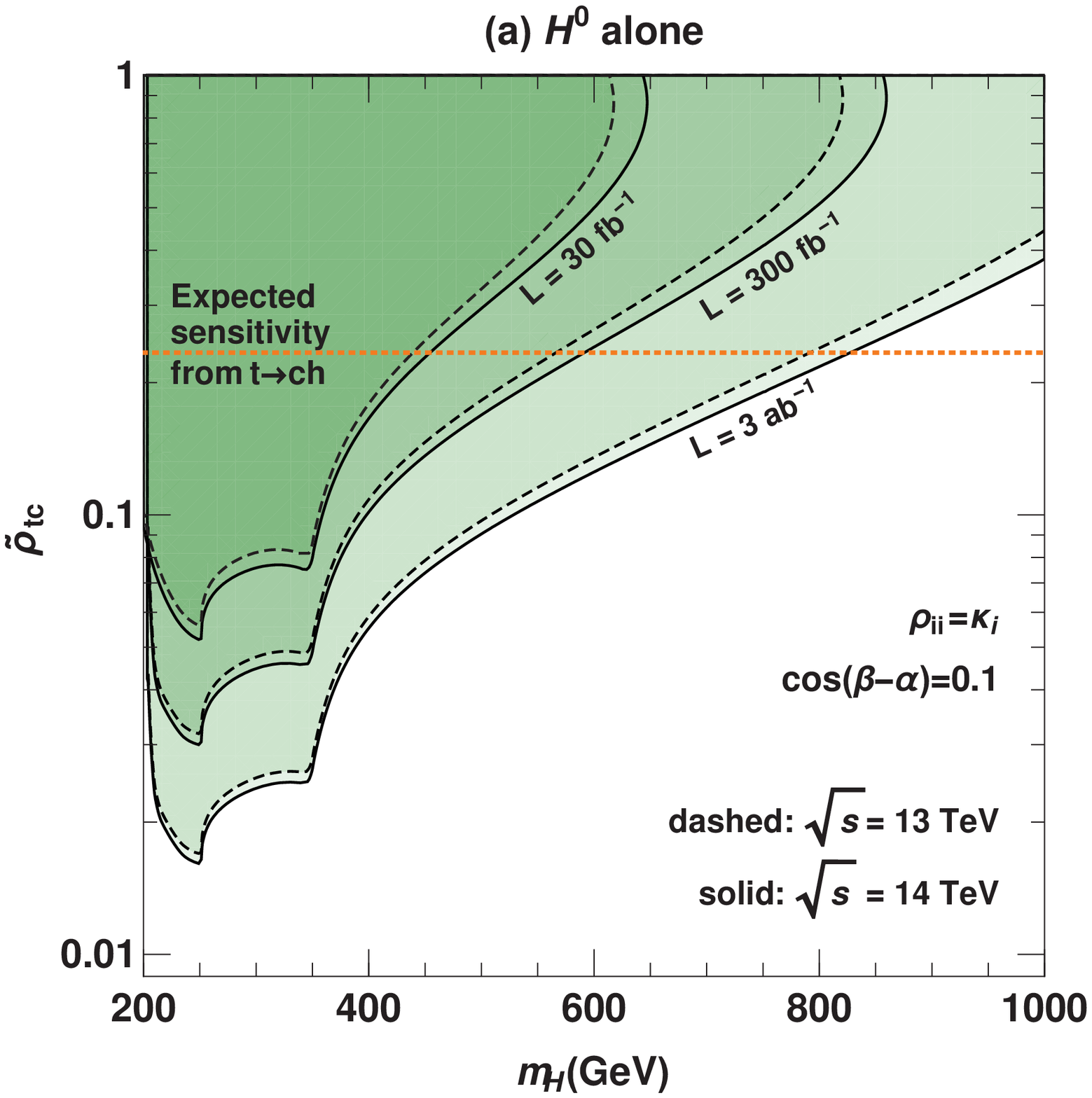} &
\includegraphics[width=66mm]{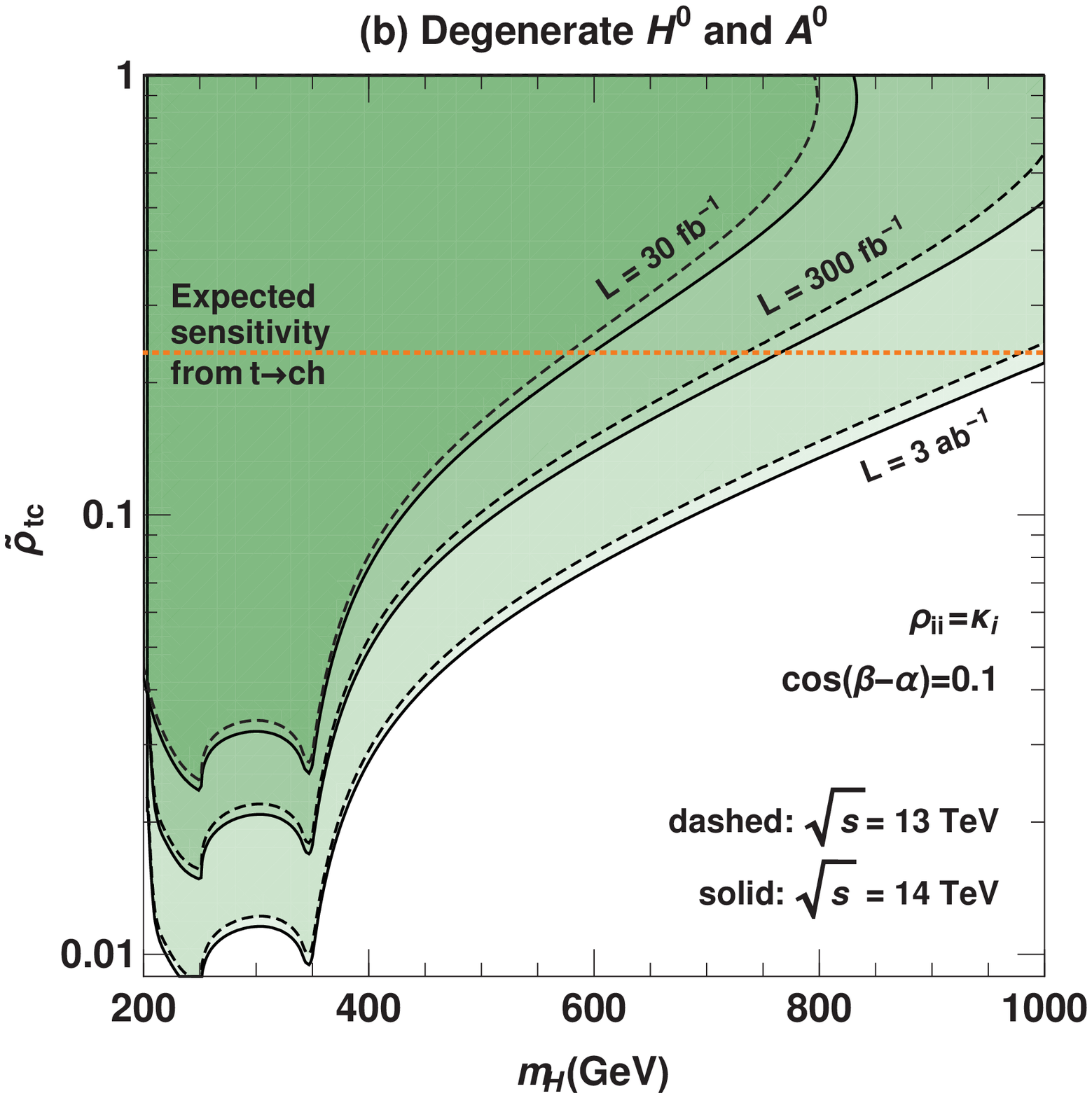} \\
\includegraphics[width=66mm]{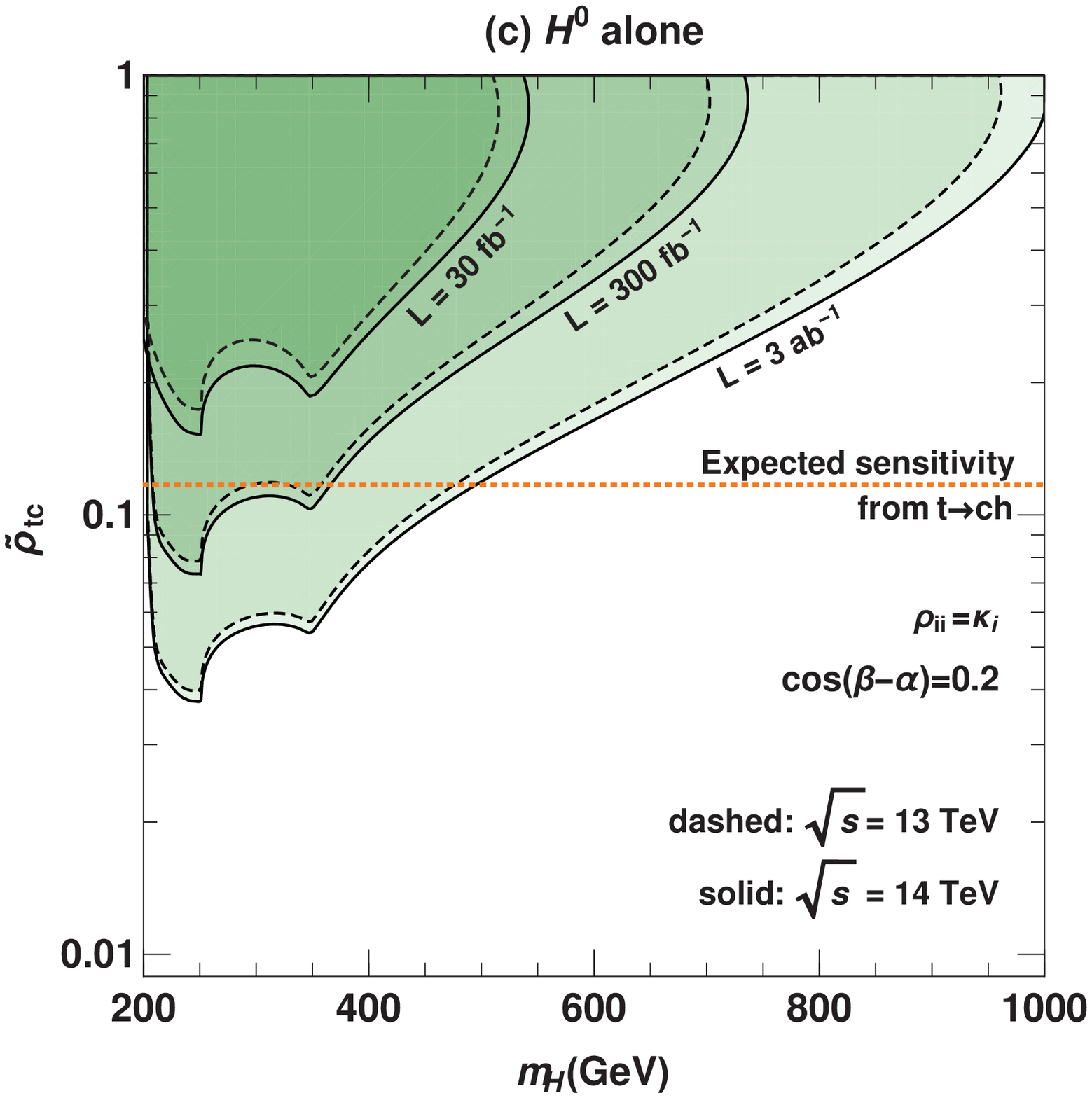} &
\includegraphics[width=66mm]{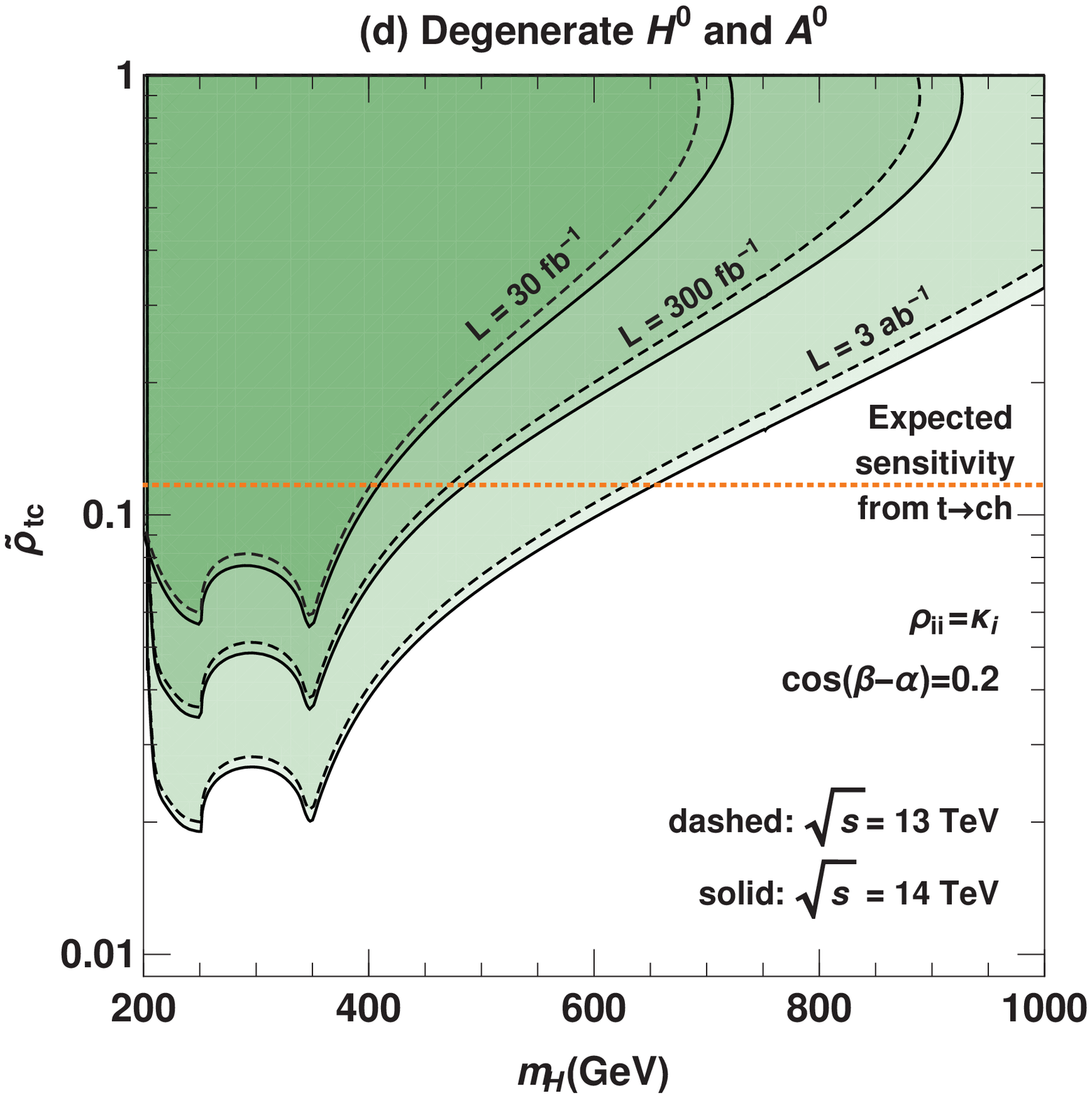}
\end{tabular}
\caption{
Discovery reach at 5$\sigma$ in the $m_\phi$--$\rho_{tc}$ plane 
for the $pp \to \phi^0 \to t\bar{c} +\bar{t}c \to bj\ell +\notE +X$ signal
at the LHC with $\sqrt{s} = 13$ (14) TeV for dashed (solid) contours.
(a) is for the heavier Higgs scalar ($H^0$) and 
(b) for the combined $H^0$ and $A^0$ assuming degeneracy,
    both for $\cos(\beta-\alpha) = 0.1$.
(c) and (d) are analogous, but for $\cos(\beta-\alpha) = 0.2$.
The discovery region is the parameter space above the contours.
Also shown is the future ATLAS sensitivity at 95 \% confidence level
for $t \to ch^0 \to c\gamma\gamma$.
\label{fig:reach}
}
\end{center}
\end{figure}

To estimate the discovery potential, we obtain the lower limit on 
$\sigma_S$ by requiring that the 99.4\%-confidence-level (CL) upper limit 
on the background is smaller than the 99.4\%-CL 
lower limit on the signal plus background~\cite{HGG}
with statistical fluctuations.
This leads to the condition,
\begin{equation}
\sigma_S \ge \frac{N}{L}\left[N+2\sqrt{L\sigma_B}\right],
\end{equation}
where $\sigma_{S\, (B)}$ is the signal (background) cross
section and $L$ the integrated luminosity.
Choosing the parameter $N = 2.5$ corresponds to 5$\sigma$ significance.
For a large number of events ($L\sigma_B \gg 1$), this requirement is
equivalent to the statistical significance
\begin{eqnarray}
N_{\rm SS} = \frac{N_S}{\sqrt{N_B}}
 = \frac{L\sigma_S}{\sqrt{L\sigma_B}} \ge 5 \, , \nonumber
\end{eqnarray}
where $N_{S\, (B)}$ is the number of signal (background) events.

We show in Fig.~\ref{fig:reach} the discovery reach
in the $m_\phi$--$\tilde{\rho}_{tc}$ plane for the FCNH signal 
$pp \to \phi^0 \to t\bar{c} +\bar{t}c \to bj\ell +\notE$ 
at the LHC with $\sqrt{s} = 13$ (14) TeV, 
for dashed (solid) contours and for $\cos(\beta-\alpha) = 0.1$ and $0.2$.
Fig.~\ref{fig:reach}(a,c) are for the heavier scalar $H^0$ alone, 
whereas Fig.~\ref{fig:reach}(b,d) are for the degenerate case, 
for which the scalar $H^0$ and pseudoscalar $A^0$ signals are added together.
The FCNH decay of the heavy Higgs will be observable for $\cbma = 0.1$
and $\tilde{\rho}_{tc} = 0.1$ up to $m_H \gtrsim 800$ GeV with 3000 fb$^{-1}$ data.
A larger value of $\rho_{tc}$ will enhance the cross section, 
hence statistical significance, of this FCNH signal.

\section{Conclusion}

In a general two Higgs doublet model, there could be
flavor changing neutral Higgs interactions with fermions.
Strong limits exist for these FCNH interactions,
except those involving the third generation quarks.
It is of great interest to study the relation between the most massive
elementary particle (the top quark) and the Higgs bosons.
The LHC has discovered a Higgs boson lighter than the top,
which makes the rare decay $t \to c h^0$ kinematically possible.
In a general 2HDM,
the decay width of $t \to c h^0$ is proportional to $\cbma$, while
that of $H^0 \to  t\bar{c}$ is proportional to $\sbma$.
Therefore, they are complementary to each other in the search for
new physics beyond the Standard Model.

We investigated the prospects for discovering $H^0,A^0 \to  t\bar{c}$ 
at the LHC, where the heavy scalar $H^0$ and pseudoscalar $A^0$ are 
produced via gluon fusion, 
which are facilitated by the extra $tt$ couplings.
The primary physics background comes from $Wjj$, $tj$, $Wbb$, $tb$,
and $t\bar{t}$.
Both signal and background processes are studied with
realistic acceptance cuts as well as tagging and mistag efficiencies.
Promising results have been found for the LHC
with a center of mass energy of 13 TeV and 14 TeV.
The FCNH decay of the heavy Higgs will be observable for $\cbma = 0.1$
and $\tilde{\rho}_{tc} = 0.1$ up to $M_H = 800$ GeV with 3000
fb$^{-1}$ of integrated luminosity.
This result is robust against a small $\cbma$,
independent of the $t \to ch^0$ search,
which becomes diminished.
If c-tagging efficiency can be improved~\cite{ATLAS_ctag},
the discovery potential of this FCNH signal will be greatly enhanced.

If $\tilde{\rho}_{tc} \agt 0.5$, $\BR(H^0 \to tc)$ can become comparable
to $B(H^0 \to t\bar{t})$ or surpass it.
Recently, it was suggested that next-to-leading order QCD and
electroweak corrections might swamp the signal of Higgs decays into
top quark pairs~\cite{Moretti:2012mq}. A recent analysis shows that 
$H^0 \to t\bar{t}$ with SM couplings can be very difficult to observe 
at the LHC~\cite{Craig:2015jba}.
In that case the FCNH decay of $H^0,A^0 \to t\bar{c}+\bar{t}c$ might
offer a promising opportunity to observe the heavier Higgs bosons.

We have not emphasized the $\tau$ lepton sector.
Recently, the CMS Collaboration reported unexpected $\tau\mu$
events~\cite{Khachatryan:2015kon} 
that might be explained by neutral Higgs boson decay~\cite{Harnik:2012pb}.
If this can be confirmed by the ATLAS collaboration in the near future,
or at LHC Run 2, it will be exciting new physics for FCNH interactions,
and $H^0,A^0 \to \tau^\pm\mu^\mp$,
unsuppressed by decoupling (i.e. small $\cbma$), 
could help discover the exotic scalars.

\section*{Acknowledgments}
CK thanks the Academia Sinica, National Taiwan University, and
National Center for Theoretical Sciences for excellent hospitality.
BA thanks Kyu Jung Bae for discussions.
The computing for this project was performed at the OU Supercomputing
Center for Education \& Research (OSCER) at the University of Oklahoma (OU).
This research was supported in part by
the U.S. Department of Energy under Grant No.~DE-FG02-13ER41979
(CK, BA, and BM);
the Academic Summit grant MOST 103-2745-M-002-001-ASP,
as well as grant NTU-EPR-103R8915 (WSH);
and grant NSC 102-2112-M-033-007-MY3 (MK).



\end{document}